\documentclass[aps,preprint,showpacs]{revtex4}
\usepackage{graphicx}

\begin{document}

\title{Non compound nucleus fission events and standard saddle-point statistical model
}

\author{S. Soheyli\footnote{Corresponding author: s.soheyli@basu.ac.ir}, and M. K. Khalili}
\affiliation{Bu-Ali Sina University, Department of Physics, Hamedan,
 Iran}
\begin{abstract}
The large body of experimental data on the fission fragments
anisotropies are analyzed in several heavy-ion induced fission
reaction systems. The entrance channel mass asymmetry parameters of
these systems put on the both sides of the Businaro-Gallone mass
asymmetry parameters. The role of the mass numbers of the projectile
and the target in the prediction of a normal or an anomalous
behavior in angular anisotropy, as well as the validity of standard
saddle-point statistical model are considered. The average
contribution of non compound nucleus fission for the systems with an
anomalous behavior in anisotropy are also determined.
\end{abstract}

\pacs{ 25.70.Jj} \maketitle
\newpage
\section{Introduction}
 During almost seven decades of researches, an immense
body of experimental data on fission processes have been
accumulated. In addition, tremendous effort has been invested on its
theoretical understanding. Nevertheless, the full understanding of
the fission process has still not been reached. The angular
distribution of fission fragments in heavy-ion induced fission
reaction is an effective probe to study the dynamics of  fission
reactions. Non compound nucleus (NCN) fission is an important area
in the field of nuclear fission. In this process,  target and
projectile come in contact forming a composite system, in which the
system reseparates before reaching to a compact compound nucleus
(CN). Due to the presence of the NCN fission events, fission
fragment anisotropies have been observed to be anomalous in
comparison to the prediction of standard saddle-point statistical
model (SSPSM), as well as the widths of the fission fragment mass
distributions have been observed to be large in comparison to the
compound nucleus fission events. In addition, the entrance channel
properties, such as the mass asymmetry ($\alpha$) of the interacting
systems with respect to the Businaro-Gallone mass asymmetry
parameter ($\alpha_{BG}$), deformation of interacting nuclei, the
bombarding energy relative to the fusion barrier, the nuclear
orientation of the interacting nuclei such as the collision with the
sides of the deformed target nucleus, and the product of $Z_PZ_T$ of
the interacting systems (where, $Z_P$ and $Z_T$ are  projectile and
target atomic numbers, respectively) play an important role in the
formation of CN.
It is also reported that with deformed targets/projectiles shell
effects play a major role in the survival probability of the CN
~\cite{Hinde1:1995, Chiskov:2003}.
It is well known that the SSPSM as a standard theory of fission
fragment angular distributions has been generally used to explain
the observed anisotropy data and it is based on the assumption that
the fission fragments are emitted along the  symmetry axis of the
fissioning nucleus and the K component of the total angular momentum
I along the symmetry axis is conserved during the descent saddle to
scission point~\cite{Wagemans:1991}. Although, the SSPSM as the
oldest model has had outstanding success for several induced fission
reactions by lighter projectiles, but the angular anisotropies for
several heavy-ion induced fusion-fission reactions are significantly
higher than those expected from the SSPSM predictions.  A majority
of the existing models attribute the observation of anomalous
behaviors in angular anisotropies of fission fragments to the
presence of NCN fission (NCNF) mechanisms such as quasi fission
(QF), fast fission (FF), and pre-equilibrium fission (PEF), rather
than to the breakdown of the SSPSM. It is reported that for the
induced reactions  by heavy projectiles $(A_{P}\geq20)$, on various
targets above the fusion barrier, the measured angular anisotropies
were larger than the SSPSM predictions~\cite{Back1:1985,Toke:1985}
as well as, these anomalous behaviors in angular anisotropy were
attributed to the contribution of NCNF ( QF ) events. Nevertheless,
we observe  normal behaviors in the measured anisotropies for  many
induced fission reactions by heavy projectiles $(A_{P}\geq20)$
[4-6]. However, later experimental angular anisotropies were
obtained for the reactions induced by light projectiles
$(A_{P}\leq20)$, on actinide targets in which the anisotropy could
not be explained by SSPSM. For example, in the
$^{16}\textrm{O}+^{238}\textrm{U}$ reaction system, the contribution
of NCNF was related to  the deformed actinide target
nucleus~\cite{Hinde1:1995, Hinde2:1996}. While, in a systematic
study on the induced fission of $^{238}\textrm{U}$ and
$^{232}\textrm{Th}$ targets by $^{16}\textrm{O}$ and
$^{19}\textrm{F}$ projectiles, as well as for
$^{14}\textrm{N}+^{232}\textrm{Th}$ reaction system at energies near
coulomb barrier, it is observed  anomalous behaviors in the fission
fragment anisotropies, it is found  normal behaviors in measured
anisotropies for several reactions induced by light projectiles
$(A_{P}\leq20)$ [4-6, 9-14]. In the literature, it is reported that
for the systems with the entrance mass asymmetry $\alpha$
$[\alpha=\frac{(A_{T}-A_{p})}{(A_{T}+A_{p})}]$ greater than
Businaro-Gallone critical mass asymmetry parameter $\alpha_{BG}$
[$\alpha_{BG}$ is parameterized as $\alpha_{BG}=0$ for
$\chi<\chi_{BG}$, and
$\alpha_{BG}=1.12\sqrt{\frac{(\chi-\chi_{BG})}{(\chi-\chi_{BG})+0.24}}$
for $\chi>\chi_{BG}$, where $\chi$  is fissility parameter, and
$\chi_{BG}=0.396$ ~\cite{Abe:1986} ] , the measured fragment
anisotropies are in agreement with the SSPSM predictions, while in
the case $\alpha<\alpha_{BG}$, the experimental fragment
anisotropies obviously  deviate from the SSPSM calculations [12,
15]. However, for $^{11}\textrm{B}+^{232}\textrm{Th}$ [9] reaction
system having $\alpha>\alpha_{BG}$, as well as for
$^{19}\textrm{F}+^{208}\textrm{Pb}$ [16],
$^{16}\textrm{O}+^{208}\textrm{Pb}$ [17],
$^{19}\textrm{F}+^{209}\textrm{Bi}$ [18] and
$^{16}\textrm{O}+^{209}\textrm{Bi}$ [19] reaction systems with
$\alpha<\alpha_{BG}$, the angular anisotropies show  anomalous and
normal behaviors, respectively.  There are several heavy ion induced
systems having $\alpha>\alpha_{BG}$ with  anomalous behaviors in
angular anisotropies,  as well as  systems having
$\alpha<\alpha_{BG}$ with  normal behaviors in angular anisotropies
as indicated in Table I.

\begin{table}[ht]
\begin{center}
\renewcommand{\arraystretch}{0.3}
\begin{tabular}{c  c c c }
\hline fission systems&~~~~Comparison between $\alpha$ and $\alpha_{BG}$&~~~~$NCNF$ contribution&~~~References\\
\hline \hline
${^{9}\textrm{Be}+^{232}\textrm{Th}}$&$\alpha(=0.925)$$>$$\alpha_{BG}(=0.882)$& Yes&~\cite{Appannababu2:2011}\\
${^{11}\textrm{B}+^{243}\textrm{Am}}$&$\alpha(=0.913)$$>$$\alpha_{BG}(=0.903)$& Yes&~\cite{Tripathi3:2007}\\
${^{12}\textrm{C}+^{232}\textrm{Th}}$&$\alpha(=0.902)$$>$$\alpha_{BG}(=0.890)$& Yes&~\cite{Thomas:2003}\\
${^{12}\textrm{C}+^{235}\textrm{U}}$&$\alpha(=0.903)$$>$$\alpha_{BG}(=0.898)$& Yes&~\cite{Thomas:2003, Lestone:1997}\\
${^{12}\textrm{C}+^{236}\textrm{U}}$&$\alpha(=0.903)$$>$$\alpha_{BG}(=0.897)$& Yes&~\cite{Thomas:2003, Lestone:1997}\\
${^{12}\textrm{C}+^{238}\textrm{U}}$&$\alpha(=0.904)$$>$$\alpha_{BG}(=0.896)$& Yes&~\cite{Thomas:2003, Lestone:1997}\\
${^{16}\textrm{O}+^{182}\textrm{W}}$&$\alpha(=0.838)$$<$$\alpha_{BG}(=0.840)$& $No$&~\cite{Hinde4:2000}\\
${^{16}\textrm{O}+^{186}\textrm{Os}}$&$\alpha(=0.842)$$<$$\alpha_{BG}(=0.850)$& $No$&~\cite{Rafiei:2008}\\
${^{16}\textrm{O}+^{188}\textrm{Os}}$&$\alpha(=0.843)$$<$$\alpha_{BG}(=0.849)$& $No$&~\cite{Rafiei:2008}\\
${^{16}\textrm{O}+^{194}\textrm{Pt}}$&$\alpha(=0.848)$$<$$\alpha_{BG}(=0.863)$& $No$&~\cite{Prasad:2010}\\
${^{16}\textrm{O}+^{197}\textrm{Au}}$&$\alpha(=0.850)$$<$$\alpha_{BG}(=0.861)$& $No$&~\cite{Appannababu:2011}\\
${^{18}\textrm{O}+^{197}\textrm{Au}}$&$\alpha(=0.833)$$<$$\alpha_{BG}(=0.860)$& $No$&~\cite{Appannababu:2009}\\
${^{19}\textrm{F}+^{184}\textrm{W}}$&$\alpha(=0.813)$$<$$\alpha_{BG}(=0.843)$& $No$&~\cite{Nath:2011}\\
${^{19}\textrm{F}+^{188}\textrm{Os}}$&$\alpha(=0.816)$$<$$\alpha_{BG}(=0.853)$& $No$&~\cite{Mahata2:2003}\\
${^{19}\textrm{F}+^{192}\textrm{Os}}$&$\alpha(=0.820)$$<$$\alpha_{BG}(=0.849)$& $No$&~\cite{Mahata2:2003}\\
${^{19}\textrm{F}+^{194}\textrm{Pt}}$&$\alpha(=0.822)$$<$$\alpha_{BG}(=0.861)$& $No$&~\cite{Mahata1:2002}\\
${^{19}\textrm{F}+^{197}\textrm{Au}}$&$\alpha(=0.824)$$<$$\alpha_{BG}(=0.865)$& $No$&~\cite{Tripathi2:2005}\\
${^{19}\textrm{F}+^{198}\textrm{Pt}}$&$\alpha(=0.825)$$<$$\alpha_{BG}(=0.858)$& $No$&~\cite{Mahata1:2002}\\
${^{24}\textrm{Mg}+^{178}\textrm{Hf}}$&$\alpha(=0.762)$$<$$\alpha_{BG}(=0.850)$& $No$&~\cite{Rafiei:2008}\\
${^{24}\textrm{Mg}+^{192}\textrm{Os}}$&$\alpha(=0.778)$$<$$\alpha_{BG}(=0.865)$& $No$&~\cite{Tripathi4:2008}\\
${^{24}\textrm{Mg}+^{197}\textrm{Au}}$&$\alpha(=0.783)$$<$$\alpha_{BG}(=0.879)$& $No$&~\cite{Tripathi4:2008}\\
${^{27}\textrm{Al}+^{186}\textrm{W}}$&$\alpha(=0.764)$$<$$\alpha_{BG}(=0.861)$& $No$&~\cite{Appannababu:2011}\\
${^{28}\textrm{Si}+^{176}\textrm{Yb}}$&$\alpha(=0.725)$$<$$\alpha_{BG}(=0.849)$& $No$&~\cite{Tripathi1:2009}\\
${^{34}\textrm{S}+^{168}\textrm{Er}}$&$\alpha(=0.663)$$<$$\alpha_{BG}(=0.850)$& $No$&~\cite{Morton:2000}\\
\hline \hline
\end{tabular}
\caption{\label{eq. table} Heavy ion induced fission systems with
unexpected behaviors in angular anisotropies of fission fragments.
These behaviors are not expected by the comparison between the
entrance channel mass asymmetry  ( $\alpha$) and the
Businaro-Gallone critical mass asymmetry ($\alpha_{BG}$).}
\end{center}
\end{table}

The model of Ramamurthy and Kapoor ~\cite{Ramamurthy:1985} gives a
quantitative estimate of the effect of NCNF on fission fragment
angular distribution. According to this model, the probability of
NCNF events ($P_{NCNF}$ ) is given by an approximate expression as
follows

\begin{equation}
P_{_{NCNF}}(I)= \exp[-0.5B_{f}(I,K=0)/T_{sad}],
\end{equation}
where, $B_{f}$ and $T_{sad}$ are the fission barrier height and the
temperature at  the saddle-point, respectively. Recently, the
investigations of the fission fragment mass angle correlations and
mass ratio distributions, as well as the analysis of the variance of
the mass distributions as a function of temperature and angular
momentum is used for the presence of QF in heavy ion induced fission
reactions ~\cite{Appannababu:2011}. A sudden change  in the standard
deviation ( a sudden increase in the standard deviation as energy
decreases to below-barrier energies ) of the fission fragments mass
distribution as a function of $E_{c.m.}/V_b$ ( where $E_{c.m.}$ and
$V_b$ are the projectile energy in center of mass and the Coulomb
barrier, respectively ), the observation of an anomalous behavior in
fission fragment anisotropies, the measurement of evaporation
residue cross section and the measurement of prescission neutron
multiplicity are known as the different probes for the presence of
PEF and QF for several heavy ion induced fission systems ~\cite{
Banerjee:2011}. It must be pointed that the product of $Z_PZ_T$ (
where $Z_P$ and $Z_T$ are the projectile and target atomic numbers,
respectively ) of the interacting systems, play an important role in
the formation of the CN. Although in the past, it is predicted that
QF occurs when $Z_PZ_T\geq1600$ ~\cite{Swiatecki:1981} but recent
results show that the onset of QF starts at a $Z_PZ_T$ value equal
to nearly 1000 and plays a dominant role at higher values of
$Z_PZ_T$ ~\cite{Rafiei:2008}. With this motivation, the purpose of
the present paper is to obtain a relation in terms of projectile and
target mass numbers by analyzing the large body of experimental data
on fission anisotropies for determination of the validity of SSPSM.

\newpage
\section{Method of Calculations}

\subsection{Standard Saddle-Point Statistical Model, and the calculation of SSPSM predictions}
According to statistical theory, fission fragments angular
distribution $(W(\theta))$ for a spin zero projectile-target
combination is given by the following
expression~\cite{Vandenbosch:1973}

\begin{equation}
W(\theta)\propto\sum_{I=0}^{\infty}\frac{(2I+1)^2T_{I}\exp[-(I+\frac{1}{2})^2\sin^2{\theta}/4K_{\circ}^2]J_{0}[i(I+\frac{1}{2})^2\sin^2
{\theta}/4K_{\circ}^2]}{(2K_{\circ}^2)^{1/2}\textrm{erf}[(I+\frac{1}{2})/(2K_{\circ}^2)^{1/2}]}.
\end{equation}
Where  $T_I$, $K_{0}^{2}$, and $J_0$ are the transmission
coefficient for fission, the variance of the K distribution ( K is
the component of the angular momentum vector ($I$) on the symmetry
axis of the fissioning nucleus ), and the zeroth order Bessel
function, respectively. The variance of the K distribution is
calculated by the following relation
\begin{equation}
K_{\circ}^{2}=\frac{\Im_{eff}T_{sad}}{\hbar^{2}},
\end{equation}

$\Im_{eff}$ and $T_{sad}$ are the effective moment of inertia and
the nuclear temperature of the compound nucleus at the saddle-point,
respectively. The nuclear temperature of the compound nucleus at the
saddle-point is given by
\begin{equation}
T_{sad}=\sqrt{\frac{E_{ex}}{\emph{a}}}=\sqrt{\frac{E_{c.m.}+Q-B_{f}-E_{R}-{\nu}E_{n}}{\emph{a}}}.
\end{equation}

In this equation, $E_{ex}$ denotes the excitation energy of the
compound nucleus at the saddle-point, while $E_{c.m.}$, $Q$,
$B_{f}$, $E_{R}$, $\nu$, and $E_{n}$ represent the center-of-mass
energy of the projectile, the $Q$ value, fission barrier height,
rotational energy of the compound nucleus, the number of pre-fission
neutrons, and the excitation energy lost due to evaporation of one
neutron from the compound nucleus prior to the system reaching to
the saddle-point. The quantity $\emph{a}$ stands for the level
density parameter at the saddle-point. The fission fragment angular
distributions are characterized by anisotropy (A), defined as the
ratio of the yield at $180^{\circ}$ or $0^{\circ}$ to that at
$90^{\circ}$ $(A=\frac{W(0^{\circ} or
180^{\circ})}{W(90^{\circ})})$. The fission anisotropy in the SSPSM
($A_{SSPSM}$) is given by an approximate
formula~\cite{Vandenbosch:1973}
\begin{equation}
A_{SSPSM}\approx1+\frac{<I^{2}>}{4K_{\circ}^{2}}.
\end{equation}

In this work, $\emph{a}$~ is taken $\frac{A_{C. N.}}{8}$ (by
accounting $\frac{A_{C. N.}}{10}$ instead of $\frac{A_{C. N.}}{8}$
in the calculations, the difference will be less than 10$\%$), as
well as $\Im_{eff}$, $B_{f}$, and $E_{R}$ are accounted by the use
of rotating finite range model (RFRM)~\cite{Sierk:1986}, while
$<I^{2}>$ quantities are taken from the literature [2, 11, 31-34].
We have used the values of the literature for $\nu$ [11, 34-38] and
$E_{n}$ is taken 10 MeV [18, 39]. In the present work, the
pre-fission neutrons are taken to be emitted before the
saddle-point, since it is not straightforward to separate
experimentally the contribution of neutrons emitted before the
saddle-point and the ones emitted after the saddle-point but before
the scission point.

\subsection{Calculation of the average contribution of NCNF  anisotropies }
In recent years, heavy-ion induced fission fragments angular
distribution measurements performed at below to above barrier
energies have generated much interest due to the failure of the
predictions of the SSPSM for heavy-ion induced fission of actinide
targets ~\cite{Kailas1:1997, Choudhry:2001, Lestone2:1997}. The
effects of entrance channel parameters such as mass asymmetry,
target deformation, and target or projectile spins on fission
fragment anisotropies have been identified in the past from
systematic study of fission fragments  angular distributions at
energies around the Coulomb barrier energies in actinide targets
~\cite{Kailas1:1997}. Non equilibrium fission ( PEF, QF, and FF )
was thought to be a probable cause of this anomaly. Almost 25 years
ago, Ramamurthy and Kapoor ~\cite{Ramamurthy:1985} proposed the
pre-equilibrium fission (PEF) model to explain the anomalous
anisotropies in several heavy-ion induced fission reaction at above
barrier energies. The main difference between CN fission and PEF is
that in the latter case the K degree of freedom is not equilibrated
but other degrees such as energy and mass-asymmetry are fully
equilibrated. Therefore the assumption of symmetric mass division is
justified in case of PEF. The K distributions of PEF will be the
product of the entrance channel K distribution and the saddle point
K distribution~\cite{Lestone:1997, Vorkapic:1995}, and the narrower
of the above two K distributions governs the fragment anisotropy.
This explains the observed larger anisotropies whenever the input K
distribution is not fully equilibrated. According to this model, the
final K distribution for fissioning nuclei is given by
$P_f(K)=P_{initial}(K)P_{saddle}(K)$, where $P_{initial}(K)$ is the
K distribution for the initial di-nuclear complex and
$P_{saddle}(K)$ is the Gaussian K distribution at the saddle-point.
On the whole, the final K distribution is governed by the initial
narrow K distribution populated in the formation phase. Following
the work of ~\cite{Lestone:1997, Vorkapic:1995}, the probability of
a fissioning nucleus having the quantum numbers I and K, when
populated from an entrance channel K-state distribution with peaks
at $\widetilde{K}$ is given by
\begin{equation}
P(J, K,
\widetilde{K})=\exp[-\frac{(K-\widetilde{K})^2}{2\sigma_{K}^2}]\times\exp[-\frac{(K\hbar)^2}{2\Im_{eff}T}]
\end{equation}
$P(J, K, \widetilde{K})$ is obtained by taking the initial K-state
distribution for each I value convoluted by a Gaussian with standard
deviation $\sigma_{K}$ and multiplied by the SSPSM K-state
distribution at fission saddle-point.

 It has been observed that at sub-barrier energies all the systems with actinide targets show anomalous
anisotropies of varying extent with respect to the SSPSM. In order
to explain such anomalous behaviors in angular anisotropies at
sub-barrier energies, a few models such as dependent QF model
~\cite{Hinde1:1995, Hinde2:1996}, pre-equilibrium model, a model
with considering the incorporation of target and projectile spins
[27, 43, 44], and the entrance channel dependent K-state model
(ECD-K)~\cite{Vorkapic:1995} have been well recognized. It is
obvious that, the prediction of SSPSM  shows the anisotropy of
compound nucleus fission (CNF) events, as well as the experimental
values of anisotropies  are due to CNF and NCNF events for the
systems having  anomalous behaviors in angular anisotropies. The
average contribution of NCNF anisotropies over the energy range of
projectile ($A_{NCN}$) for these systems is given by
\begin{equation}
A_{NCN}=\frac{A_{exp}-A_{SSPSM}}{A_{exp}}.
\end{equation}

In this equation, $A_{SSPSM}$ is the average contribution of SSPSM
prediction  over the energy range of the projectile, as well as
$A_{exp}$  stands the average  experimental value of anisotropy over
the same energy range.

\section{Results and discussion}
The experimental fission fragment angular anisotropies ( A ) along
with the SSPSM predictions for the induced fission of
$^{208}\textrm{Pb}$ target by different projectiles
($^{24}\textrm{Mg}$, $^{28}\textrm{Si}$, and $^{32}\textrm{S}$) are
shown in Fig. 1. Calculation of the average of NCNF contributions
for these three systems over the
$1.05\leq\frac{E_{c.m.}}{V_{b}}\leq1.35$ energy range show that the
NCNF contributions will increase as the mass number of the
projectile increases.
\begin{figure}[ht]
\centering
\includegraphics[scale=0.7]{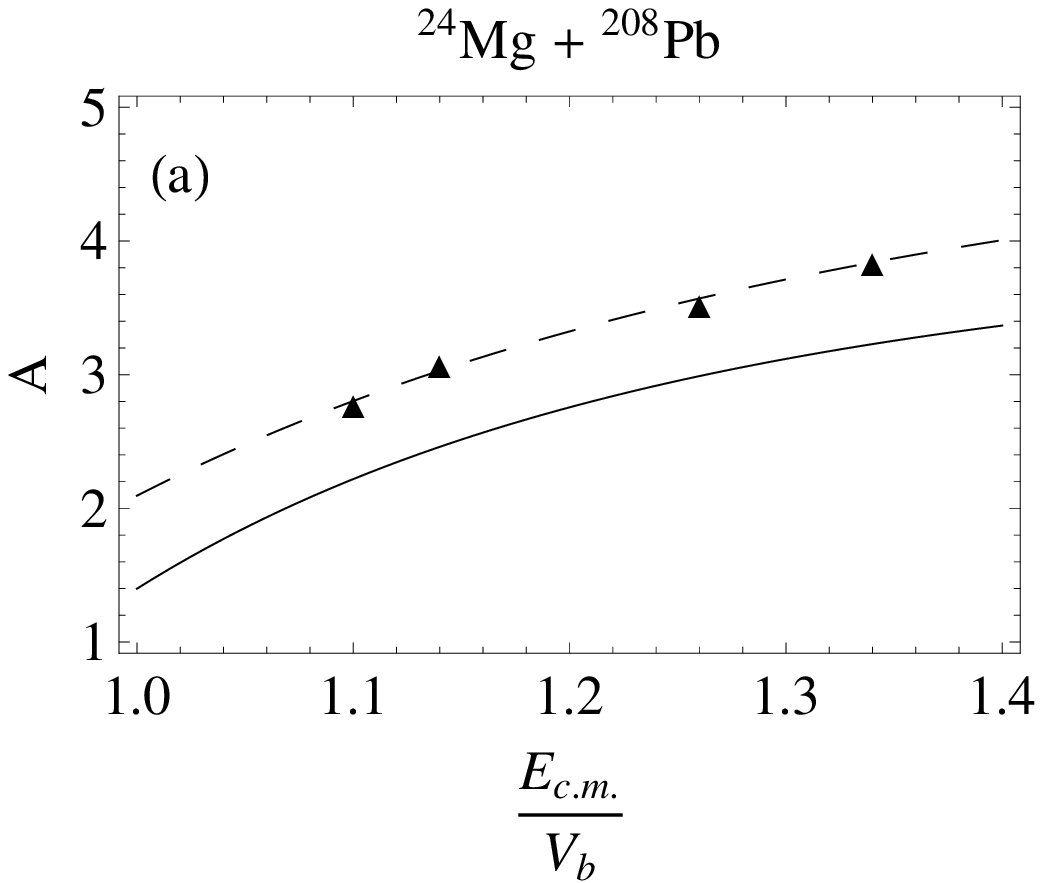}
\includegraphics[scale=0.7]{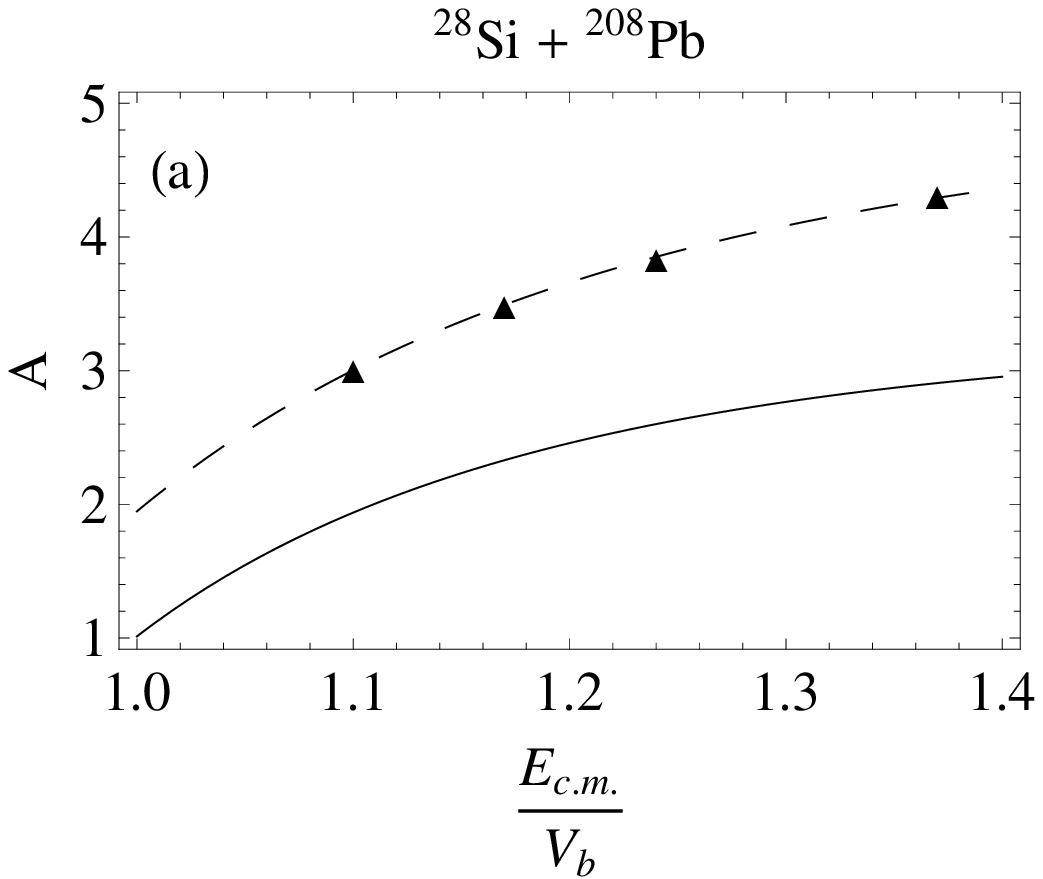}
\includegraphics[scale=0.7]{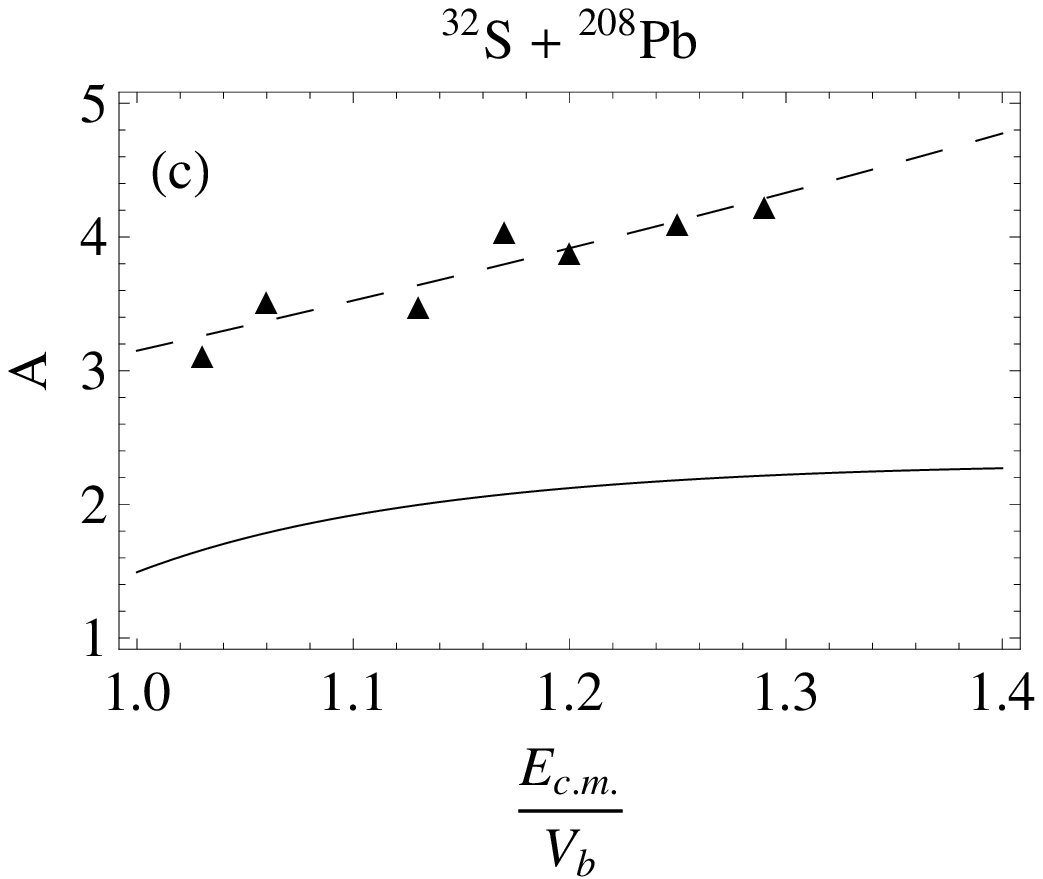}

\caption{ Experimental data of anisotropy (A) for the fission of
$^{208}\textrm{Pb}$ target induced by  several
 projectiles. (a)  Solid and dashed curves are the
SSPSM prediction and experimental value of anisotropy for the
fission of $^{24}\textrm{Mg}+^{208}\textrm{Pb}$ reaction system, (b)
Solid and dashed curves are the SSPSM prediction and experimental
value of anisotropy for the fission of
$^{28}\textrm{Si}+^{208}\textrm{Pb}$ reaction system, and (c) Solid
and dashed curves are the SSPSM prediction and experimental value of
anisotropy for the fission of $^{32}\textrm{S}+^{208}\textrm{Pb}$
reaction system~\cite{Back1:1985}.}  \label{Pb}
\end{figure}
 We also considered the induced fission of $^{238}\textrm{U}$ target
by $^{16}\textrm{O}$, $^{19}\textrm{F}$, $^{27}\textrm{Al}$
projectiles,  as well as the induced fission $^{232}\textrm{Th}$ by
$^{16}\textrm{O}$, $^{19}\textrm{F}$, and $^{32}\textrm{S}$
projectiles. The average contributions of NCNF for the induced
fission of $^{208}\textrm{Pb}$, $^{238}\textrm{U}$, and
$^{232}\textrm{Th}$ targets by different projectiles are shown in
Fig. 2. These average contributions  for the induced fission of
$^{238}\textrm{U}$, and $^{232}\textrm{Th}$ targets by different
projectiles are calculated over the
$1.0\leq\frac{E_{c.m.}}{V_{b}}\leq1.2$ and
$0.95\leq\frac{E_{c.m.}}{V_{b}}\leq1.15$  energy ranges,
respectively. In this figure, the thick solid line shows the
contributions of NCNF for induced fission of $^{208}\textrm{Pb}$
target by different projectiles. The points on this thick solid line
are  the average contributions of NCNF for the $^{24}\textrm{Mg}$, $
^{28}\textrm{Si}$, and $^{32}\textrm{S}+^{208}\textrm{Pb}$ reaction
systems. It can be observed that in induced fission of
$^{208}\textrm{Pb}$ nucleus, the contributions of NCNF for
projectiles whose the mass number is 20 or less is zero, as well as
there is the average contribution of NCNF for induced fission of
this target by projectiles with the mass number more than 20. This
result are found to be in good agreement with
experiments~\cite{Back1:1985}. In this figure,  the thin solid and
dashed lines  show the average contributions of NCNF for induced
fission of $^{238}\textrm{U}$ target by different projectiles. The
discrepancy  on these two lines is because of the different values
of $<I^2>$ for induced fission of $^{238}\textrm{\textrm{U}}$ by
$^{16}\textrm{O}$ projectile that was taken from different
references (Back \emph{et al}., [2], and Nasirov \emph{et al}.,
[34]).

\begin{figure}[h]
\centering
\includegraphics[scale=0.73]{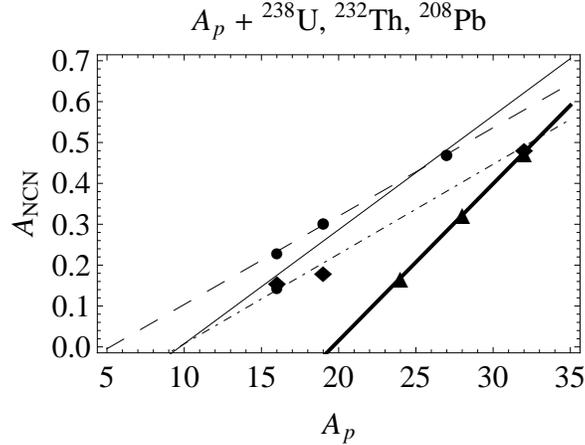}

\caption{ The average contributions of NCNF $(A_{NCN})$ versus the
mass number of projectile. Thick solid line for $^{208}\textrm{Pb}$,
thin solid and dashed lines for $^{238}\textrm{U}$ nucleus by using
Back data and by using Nasirov data, respectively, and dashed-dotted
line for $^{232}\textrm{Th}$ target. }
\end{figure}
Dashed-dotted line in the Fig. 2 shows, the average contributions of
NCNF for induced fission of $^{232}\textrm{Th}$ target by different
projectiles over the $0.95\leq\frac{E_{c.m.}}{V_{b}}\leq1.15$ energy
range. It is interesting to note that whatever the target is heavier
, one can infer the onset of  NCNF  events occurs  with lighter
projectiles and vise versa. By considering several heavy-ion induced
fission systems with anomalous behaviors in angular anisotropies,
such as $^{24}\textrm{Mg}, ^{28}\textrm{Si}$, and
$^{32}\textrm{S}+^{208}\textrm{Pb}$ in comparison with
$^{16}\textrm{O}$, $^{19}\textrm{F}$, and
$^{27}\textrm{Al}+^{238}\textrm{U}$ systems, it can be observed that
the contributions of NCNF in induced fission of heavier targets are
more than the contributions of NCNF in induced fission of lighter
target by the same projectile. While the calculated value of the
average contributions of NCNF for the
$^{32}\textrm{S}+^{197}\textrm{Au}$ system over the
$1.1\leq\frac{E_{c.m.}}{V_{b}}\leq1.2$ energy range is approximately
44\%, this contribution for the $^{32}\textrm{S}+^{208}\textrm{Pb}$
system over the same energy range is obtained 50\%.
  The predicted value of average contributions of NCNF for the $^{40}\textrm{Ar}+^{208}\textrm{Pb}$ system over the
$1.05\leq\frac{E_{c.m.}}{V_{b}}\leq1.3$ energy range from  Fig. 2 is
approximately 87\% where is a good agreement with the work of Keller
\emph{et al}.~\cite{Keller: 1987}. Itkis \emph{et al.}, measured the
mass and energy distributions of the
$^{56}\textrm{Fe}+^{208}\textrm{Pb}$ reaction system, as well as
they reported that for this reaction, the QF process dominates at
all measured energy~\cite{Itkis:2011}. This result is predicted very
well form Fig. 2.  In order to make a comparison between the average
contributions of NCNF in induced fission of different targets by the
same projectile, we have calculated these contributions for the
induced fission of $^{184}\textrm{W}$, $^{197}\textrm{Au}$, and
$^{208}\textrm{Pb}$ targets by $^{32}\textrm{S}$ projectile, as well
as those of the induced fission of $^{232}\textrm{Th}$,
$^{238}\textrm{U}$, and $^{248}\textrm{Cm}$ targets by
$^{16}\textrm{O}$ projectile over the same energy range. These
calculated contributions are shown in Fig. 3. It can be seen from
this figure that  the average contributions of NCNF for induced
fission of different targets by the same projectile begin from a
given target mass number, as well as this contributions also show a
linear behavior in terms of the mass numbers of targets for a given
projectile.

\begin{figure}[h]
\centering
\includegraphics[scale=.73]{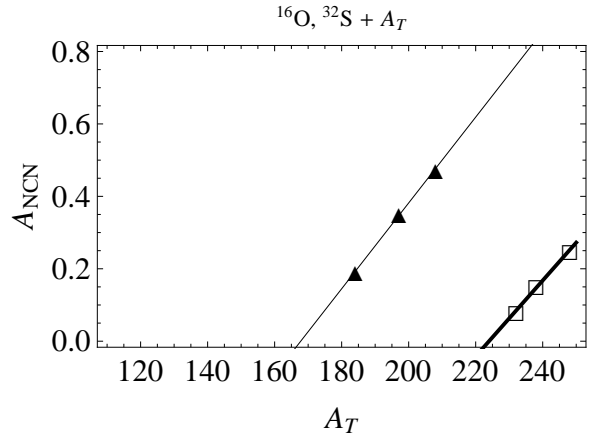}

\caption { The average contributions of NCNF $(A_{NCN})$ for induced
fission of different targets by the same projectiles. Thick solid
and thin solid lines shows $A_{NCN}$ for induced fission of
different targets by $^{16}\textrm{O}$ and $^{32}\textrm{S}$
projectiles, respectively.}
\end{figure}

Finally, as indicated in Fig. 4, for  the heavy-ion induced fission
reactions systems in which the mass numbers of target and
projectile( $A_T$, and $A_P$, respectively ) locate below the curve
shown in this figure, fission fragment angular anisotropies exhibit
normal behaviors, in addition, the predictions of SSPSM are in
agreement with the experimental angular anisotropies data. However,
for the reactions in which the $A_T$ and $A_P$ lying above this
curve, there exist an admixture of CNF and NCNF events, so that the
measured fission fragment anisotropies are anomalously large
compared to the peredictions based on the SSPSM, as well as this
model is not valid.

\begin{figure}[h]
\centering
\includegraphics[scale=0.73]{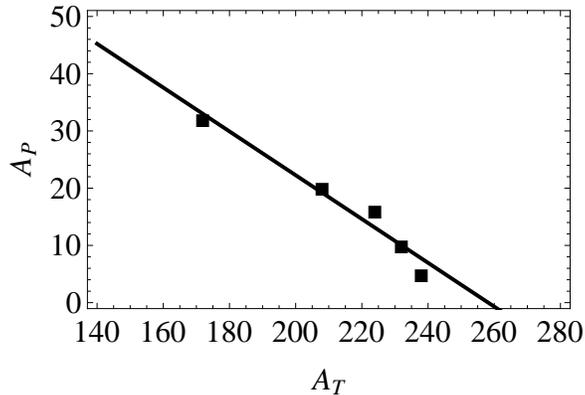}

\caption {The border diagram between compound nucleus fissions with
 compound and non-compound fissions.}
\end{figure}
It is interesting to note that Berriman \emph{et
al.,}~\cite{Berriman:2001} reported a contribution of NCNF for a
very asymmetric reaction $^{19}\textrm{F}+^{197}\textrm{Au}$  system
by measuring of the width of the fission fragments mass
distribution. However, a recent measurement of fission fragments
angular distributions for the same reaction showed no evidence of
NCNF ~\cite{Tripathi2:2005, Appannababu:2009} as can be also seen
from Fig. 4. In another work, evidence of NCNF has been found in the
$^{34}\textrm{S}+^{168}\textrm{Er}$ reaction system by considering
of fission fragment mass distribution, but no evidence for NCNF was
observed in the investigation of fission fragment angular
distributions ~\cite{Morton:2000, Rafiei:2008}. Fig. 4 shows that
the contribution of NCNF is not significant for induced fission of
$^{168}\textrm{Er}$ target by the projectiles with $A_P\leq35$. Our
results is in agreement with the experimental observations which
have been obtained up to now [4-7, 9, 11, 16-21, 23, 27, 46, 47].

\section{Summary and Conclusions}
The average contributions of NCNF have been calculated for several
heavy-ion induced fission reaction systems with having an anomalous
behaviors in fission fragments angular anisotropies by comparison
between the experimental data of  fission fragment angular
distrbutions and the predictions of SSPSM. Although, it is reported
that for the systems with $\alpha>\alpha_{BG}$, the measured fission
fragments anisotropies are in general agreement with  the
expectation of the SSPSM, as well as for the systems with
$\alpha<\alpha_{BG}$, the experimental fragment anisotropies are
considerably greater than the predictions of the SSPSM at sub
barrier and near barrier energies. However, there are the reaction
systems on the different sides of Businaro-Gallone critical mass
asymmetry with a unexpected behaviors in fission fragments angular
anisotropies as indicated in Table I. Our calculated NCNF
contributions for the reaction systems with anomalous behaviors in
angular anisotropies show that these contributions increase with
increasing the mass number of projectiles for a given target, as
well as this contribution exhibits a linear behavior as a function
of the mass number of targets for a given projectile. Finally, the
validity of SSPSM in the prediction of angular anisotropies for the
reaction systems with normal behaviors in fission fragment angular
anisotropies is also determined.

\end{document}